\documentclass[12pt,twoside, a4paper]{article}
\def\pd{\partial}
\def\mc{\mathcal}

\usepackage[dvips]{graphicx}
\usepackage{amssymb}
\usepackage{amssymb,amsmath}
\usepackage{graphicx}
\usepackage{dsfont}
\usepackage{caption}
\usepackage{subcaption}
\input{epsf.sty}  
\begin{document}
\begin{center}
\LARGE{\textbf{Supersymmetric solutions from matter-coupled $7D$ $N=2$ gauged supergravity}}
\end{center}
\vspace{1 cm}
\begin{center}
\large{\textbf{Parinya Karndumri}$^a$ and \textbf{Patharadanai Nuchino}$^b$}
\end{center}
\begin{center}
String Theory and Supergravity Group, Department
of Physics, Faculty of Science, Chulalongkorn University, 254 Phayathai Road, Pathumwan, Bangkok 10330, Thailand
\end{center}
E-mail: $^a$parinya.ka@hotmail.com \\
E-mail: $^b$danai.nuchino@hotmail.com \vspace{1 cm}\\
\begin{abstract}
We study supersymmetric solutions within seven-dimensional $N=2$ gauged supergravity coupled to three vector multiplets in seven dimensions. The gauged supergravity contains six vector fields that gauge the $SO(4)\sim SO(3)\times SO(3)$ symmetry and admits two $N=2$ supersymmetric $AdS_7$ vacua with $SO(4)$ and $SO(3)_{\textrm{diag}}\subset SO(3)\times SO(3)$ symmetries. We consider solutions interpolating between two asymptotically locally $AdS_7$ geometries in the presence of a three-form field. For a particular value of the two $SO(3)$ gauge coupling constants, the $SO(3)_{\textrm{diag}}$ supersymmetric $AdS_7$ vacuum does not exist, but the solutions can be uplifted to eleven dimensions by a known reduction ansatz. We also study solutions of this type and their embedding in M-theory. We further extend these solutions to include the $SO(3)_{\textrm{diag}}$ gauge fields and argue that, in general, this generalization does not lead to supersymmetric solutions.             
\end{abstract}
\newpage
\section{Introduction}
Over the past twenty years, the AdS/CFT correspondence has been widely tested and confirmed by a large number of interesting results. It has been applied to holographic studies of strongly coupled field theories in various space-time dimensions. One of the interesting cases is AdS$_7$/CFT$_6$ correspondence which has been argued to describe the dynamics of M$5$-branes in M-theory since the first proposal of the correspondence in \cite{maldacena}.   
\\
\indent As in other cases, AdS$_7$/CFT$_6$ correspondence can be efficiently investigated by using gauged supergravities in seven dimensions. For example, $AdS_7\times S^4$ geometry of M-theory, dual to $N=(2,0)$ superconformal field theory (SCFT) in six dimensions, can be described by $SO(5)$ $N=4$ gauged supergravity \cite{11D_to_7D_Townsend,11D_to_7D_Nastase}. In this paper, we are interested in the case of half-maximal $N=2$ gauged supergravity. The corresponding $AdS_7$ vacua are dual to $N=(1,0)$ SCFTs, see for example \cite{Berkooz_6D_dual,Ferrara_AdS7CFT6}. Some of these SCFTs can be obtained from an orbifold of the $N=(2,0)$ SCFTs \cite{AdS7_orbifold1,AdS7_orbifold2}, and recently, an interest in $N=(1,0)$ SCFTs has been increased by many new results, see \cite{All_AdS7,N1_0_Gaiotto,gauge_gravity_6D_7D,new_6D_fixed_point,AdS7_CFT6_orientifold} for an incomplete list. 
\\
\indent $N=2$ gauged supergravity has been constructed for a long time in \cite{Eric_N2_7D,Park_7D,Salam_7DN2}. These theories, however, do not admit any $AdS_7$ vacua. The existence of an $AdS_7$ vacuum requires an additional deformation in the form of a mass term for the three-form field, dual to the two-form field in the $N=2$ gravity multiplet. The $N=2$ gauged supergravity including both types of deformations has been given in \cite{Pure_N2_7D1}, see also \cite{Pure_N2_7D2}. An extension of this $N=2$ gauged supergravity to include vector multiplets has been given in \cite{Eric_N2_7Dmassive}. A number of supersymmetric $AdS_7$ vacua and various types of holographic solutions within this gauged supergravity have been studied in \cite{7D_flow,7D_noncompact,7D_twist}. A classification of possible gauge groups that can give rise to maximally supersymmetric $AdS_7$ vacua has also been given in \cite{AdS_7_N2_Jan}. Most of the previously known solutions of pure and matter-coupled $N=2$ gauged supergravity only involve the metric and scalar fields although the results of \cite{7D_noncompact} and \cite{7D_twist} do include solutions with non-vanishing gauge fields.   
\\
\indent Supersymmetric solutions of pure $N=2$ gauged supergravity with all bosonic fields, including the three-form and gauge fields, non-vanishing have appeared recently in \cite{7D_sol_Dibitetto} along with the embedding in M-theory by using the result of \cite{Pope_N2_7D}. The solution without the $SU(2)$ gauge fields has also been uplifted to massive type IIA theory in \cite{6D_surface_Dibitetto} in which the solution is interpreted as a two-dimensional conformal defect in $N=(1,0)$ SCFT. In the present work, we are interested in similar solutions in $N=2$ gauged supergravity coupled to three vector multiplets with $SO(4)\sim SO(3)\times SO(3)$ gauge group. In this case, the maximally supersymmetric $AdS_7$ vacuum is dual to an $N=(1,0)$ SCFT with flavor symmetry $SO(3)$. The solutions presented here should give an extension of the results in \cite{7D_flow} and \cite{7D_sol_Dibitetto} and represent more general solutions of $N=2$ seven-dimensional gauged supergravity. 
\\
\indent 
The paper is organized as follow. In section \ref{N2_7D_SUGRA}, we give a short review of the
matter coupled $N=2$ gauged supergravity in seven dimensions. The aim of this section is to give
relevant formulae which will be used throughout the paper. Supersymmetric solutions interpolating between two asymptotically locally $AdS_7$ geometries and solutions flowing from an asymptotically locally $AdS_7$ with $SO(4)$ symmetry to a singular geometry are given in section \ref{DW_3form}. These solutions are obtained by using the $AdS_3\times S^3$-sliced domain wall ansatz. We also discuss the embedding of the latter type of solutions in eleven-dimensional supergravity. In section \ref{DW_3form_vector}, we
study similar solutions with non-vanishing $SO(3)_{\textrm{diag}}$ gauge fields and argue that this does not give rise to supersymmetric solutions. Some conclusions and comments on
the results are given in section \ref{conclusion}. All bosonic field equations of $N=2$ gauged supergravity coupled to vector multiplets are given in appendix \ref{field_equations}. A consistent reduction ansatz for special values of the gauge coupling constants is reviewed in appendix \ref{reduction_ansatz}.

\section{$N=2$ gauged supergravity in seven dimensions}\label{N2_7D_SUGRA}
We first give a brief review of $N=2$ gauged
supergravity in seven dimensions with topological mass term. All of
the conventions and notations are essentially the same as those in \cite{Eric_N2_7Dmassive} to
which the reader is referred for more detail. 
\\
\indent The half-maximal $N=2$ supergravity in seven dimensions can couple to an arbitrary number $n$
of vector multiplets, the only matter multiplets in $N=2$ supersymmetry. The field
contents are given respectively by
\begin{eqnarray}
\textrm{Supergravity multiplet}&:&\qquad (e^{\hat{\mu}}_\mu, \psi^A_\mu,
A^i_\mu,\chi^A,B_{\mu\nu},\sigma) \nonumber \\
\textrm{Vector multiplet}&:&\qquad (A_\mu,\lambda^A,\phi^i)^r\, .
\end{eqnarray}
Curved and flat space-time indices are denoted by $\mu,\nu,\ldots$ and
$\hat{\mu},\hat{\nu},\ldots$, respectively. $B_{\mu\nu}$ and $\sigma$ are the two-form
and the dilaton fields. The two-form field will be dualized to a
three-form field $C_{\mu\nu\rho}$ which admits a topological mass term leading to a massive deformation of the $N=2$ supergravity. Indices $i,j=1,2,3$ label triplets of $SU(2)_R\sim SO(3)_R$ symmetry. The dilaton $\sigma$ can be described by a coset space $SO(1,1)\sim \mathbb{R}^+$. 
\\
\indent Each of the vector multiplets, labelled by indices $r,s=1,2,\ldots , n$, consists of a vector field $A_\mu$, two gaugini $\lambda^A$ and $3$ scalars $\phi^i$. Indices $A,B,\ldots =1,2$ label a doublet of the $SU(2)_R$ symmetry and will be generally suppressed throughout this paper. There are $3n$ scalar fields $\phi^{ir}$ parametrizing
$SO(3,n)/SO(3)\times SO(n)$ coset manifold. These can be efficiently described by a coset representative of the form
\begin{equation}
L=(L_I^{\phantom{I}i},L_I^{\phantom{I}r}), \qquad I=1,\ldots, 3+n\, .
\end{equation}
The inverse of $L$ will be denoted by
\begin{equation}
L^{-1}=(L^I_{\phantom{I}i},L^I_{\phantom{I}r}). 
\end{equation}
\indent Since $L$ is an element of $SO(3,n)$, we have the following relations 
\begin{eqnarray}
\eta_{IJ}&=&-L_I^{\phantom{I}i}{L_J}^i+L_I^{\phantom{I}r}L_J^{\phantom{J}r},\nonumber \\
L^I_{\phantom{I}i}&=&\eta^{IJ}L_{Ji}\qquad \textrm{and}\qquad L^I_{\phantom{I}r}=\eta^{IJ}L_{Jr}. 
\end{eqnarray}
It should be noted that indices $i,j$ and $r,s$ are raised and lowered by $\delta_{ij}$ and $\delta_{rs}$, respectively
while the full $SO(3,n)$ indices $I,J$ are raised and lowered by the $SO(3,n)$ invariant tensor $\eta_{IJ}=\textrm{diag}(---+\ldots +)$. With these conventions, relations involving components of $L$ can be written as
\begin{equation}
{L_I}^i=L_{Ii},\qquad {L_I}^r=L_{Ir},\qquad 
{L_I}^iL^I_{\phantom{I}j}=-\delta^i_j,\qquad
{L_I}^r{L^I}_s=\delta^{r}_s\, .
\end{equation}
\indent Gaugings of $N=2$ supergravity can be obtained by promoting a subgroup $G_0$ of the global symmetry group $\mathbb{R}^+\times SO(3,n)$ to be a local symmetry. If the gauging does not invove the $\mathbb{R}^+$ factor, the embedding of $G_0$ in $SO(3,n)$ is described by the $SO(3,n)$ tensor ${f_{IJ}}^K$ identified with the structure constants of the gauge group $G_0$ via the gauge algebra
\begin{equation}
[T_I,T_J]={f_{IJ}}^KT_K
\end{equation}
where $T_I$ denote the gauge generators. In the embedding tensor formalism, ${f_{IJ}}^K$ is a component of the full embedding tensor, see \cite{7D_embedding_tensor_Dibitetto} for more detail.  
\\
\indent For the gauging to be a consistent one, preserving all of the original supersymmetry, ${f_{IJ}}^K$ must satisfy the conditions
\begin{equation}
f_{IJK}=\eta_{KL}{f_{IJ}}^L=f_{[IJK]}\qquad \textrm{and}\qquad {f_{[IJ}}^L{f_{K]L}}^M=0\, .
\end{equation}
Apart from the gauging, there is also a massive deformation given by adding a topological mass term to the three-form field $C_{\mu\nu\rho}$. This additional deformation is crucial for the gauged supergravity to admit $AdS_7$ vacua.  
\\
\indent The bosonic Lagrangian including both the gauging and the massive deformation can be written as
\begin{eqnarray}
\mc{L}&=&\frac{1}{2}R*\mathbb{I}-\frac{1}{2}e^\sigma a_{IJ}*F^I_{(2)}\wedge F^J_{(2)}-\frac{1}{2}e^{-2\sigma}*H_{(4)}\wedge H_{(4)}-\frac{5}{8}*d \sigma \wedge d\sigma\nonumber \\
& &-\frac{1}{2}*P^{ ir}\wedge P_{ir}+\frac{1}{\sqrt{2}}H_{(4)}\wedge
\omega_{(3)}-4hH_{(4)}\wedge C_{(3)}-V*\mathbb{I}\label{7D_action}
\end{eqnarray}
where we have used the form language for convenience in dealing with the field equations. The constant $h$ describes the topological mass term for the three-form $C_{(3)}$ with $H_{(4)}=dC_{(3)}$. 
\\
\indent The scalar potential is given by
\begin{equation}
V=\frac{1}{4}e^{-\sigma}\left(C^{ir}C_{ir}-\frac{1}{9}C^2\right)+16h^2e^{4\sigma}
-\frac{4\sqrt{2}}{3}he^{\frac{3\sigma}{2}}C
\end{equation}
where $C$ and $C_{ir}$ are defined in term of the coset representative as
\begin{equation}
C=-\frac{1}{\sqrt{2}}f_{IJ}^{\phantom{sad}K}L^I_{\phantom{s}i}L^J_{\phantom{s}j}L_{Kk}\epsilon^{ijk},\qquad C_{ir}=\frac{1}{\sqrt{2}}f_{IJ}^{\phantom{sad}K}L^I_{\phantom{s}j}L^J_{\phantom{s}k}L_{Kr}\epsilon^{ijk}\, .
\end{equation}
\indent The scalar kinetic term is written in term of the vielbein $P^{ir}_\mu$ on the $SO(3,n)/SO(3)\times SO(n)$ coset defined by
\begin{equation}
P_\mu^{ir}=L^{Ir}\left(\delta^K_I\pd_\mu+{f_{IJ}}^{K}A_\mu^J\right){L_K}^i\, .
\end{equation}
The scalar matrix $a_{IJ}$ appearing in the kinetic term of vector fields is given by 
\begin{equation}
a_{IJ}={L_I}^iL_{Ji}+{L_I}^rL_{Jr}\, .
\end{equation}
Finally, the Chern-Simons three-form satisfying $d\omega_{(3)}=F^I_{(2)}\wedge F^I_{(2)}$ is defined by
\begin{equation}
\omega_{(3)}=F^I_{(2)}\wedge
A^I_{(1)}-\frac{1}{6}f_{IJ}^{\phantom{sa}K}A^I_{(1)}\wedge
A^J_{(1)}\wedge A_{(1)K}
\end{equation}
with the gauge field strength tensors $F^I_{(2)}=dA^I_{(1)}+\frac{1}{2}{f_{JK}}^{I}A^J_{(1)}\wedge A^K_{(1)}$. The associated bosonic field equations are collected in appendix \ref{field_equations}. The gauge coupling constants are included in ${f_{IJ}}^K$.
\\
\indent Other ingredients which are relevant for finding supersymmetric solutions are supersymmetry transformations of fermions. With all fermionic fields vanishing, these are given by
\begin{eqnarray}
\delta \psi_\mu &=&2D_\mu
\epsilon-\frac{\sqrt{2}}{30}e^{-\frac{\sigma}{2}}C\gamma_\mu
\epsilon
-\frac{1}{240\sqrt{2}}e^{-\sigma}H_{\rho\sigma\lambda\tau}\left(\gamma_\mu
\gamma^{\rho\sigma\lambda
\tau}+5\gamma^{\rho\sigma\lambda\tau}\gamma_\mu\right)\epsilon\nonumber
\\
&
&-\frac{i}{20}e^{\frac{\sigma}{2}}F^i_{\rho\sigma}\sigma^i\left(3\gamma_\mu
\gamma^{\rho\sigma}-5\gamma^{\rho\sigma}\gamma_\mu\right)\epsilon
-\frac{4}{5}he^{2\sigma}\gamma_\mu \epsilon,\label{delta_psi}\\
\delta \chi &=&-\frac{1}{2}\gamma^\mu\pd_\mu \sigma
\epsilon-\frac{i}{10}e^{\frac{\sigma}{2}}F^i_{\mu\nu}\sigma^i\gamma^{\mu\nu}\epsilon-
\frac{1}{60\sqrt{2}}e^{-\sigma}H_{\mu\nu\rho\sigma}\gamma^{\mu\nu\rho\sigma}\epsilon\nonumber
\\
& &+\frac{\sqrt{2}}{30}e^{-\frac{\sigma}{2}}C\epsilon-\frac{16}{5}e^{2\sigma}h\epsilon,\label{delta_chi}\\
\delta \lambda^r &=&i\gamma^\mu
P^{ir}_\mu\sigma^i\epsilon-\frac{1}{2}e^{\frac{\sigma}{2}}F^r_{\mu\nu}\gamma^{\mu\nu}\epsilon-\frac{i}{\sqrt{2}}e^{-\frac{\sigma}{2}}C^{ir}\sigma^i\epsilon\label{delta_lambda}
\end{eqnarray}
where $\sigma^i$ are the usual Pauli matrices. The covariant derivative of $\epsilon$ is given by
\begin{equation}
D_\mu\epsilon=\pd_\mu \epsilon+\frac{1}{4}{\omega_\mu}^{ab}\gamma_{ab}\epsilon+\frac{1}{2\sqrt{2}}Q^i_\mu \sigma^i\epsilon
\end{equation}
where $Q^i_\mu=\frac{i}{\sqrt{2}}\epsilon^{ijk}Q^{jk}_{\mu}$ is defined in term of the composite connection
\begin{equation}
Q^{ij}_\mu=L^{Ij}\left(\delta^K_I\pd_\mu +{f_{IJ}}^KA^J_\mu\right){L_K}^i\, .
\end{equation}

\section{$AdS_3\times S^3$-sliced domain walls with the three-form field}\label{DW_3form}
In this section, we will study supersymmetric solutions involving the seven-dimensional metric, scalars and the three-form field. We will consider the case of $n=3$ vector multiplets and $SO(4)\sim SO(3)\times SO(3)$ gauge group. The first $SO(3)$ factor is identified with the $SO(3)_R\sim SU(2)_R$ R-symmetry. The corresponding structure constants are given by
\begin{equation}
f_{IJK}=(g_1\epsilon_{ijk},g_2 \epsilon_{rst}).
\end{equation}
For a particular case of $g_2=g_1$, the resulting gauged supergravity can be embedded in eleven-dimensional supergravity \cite{7D_from_11D}. 
\\
\indent An explicit parametrization of $SO(3,3)/SO(3)\times SO(3)$ coset can be achieved by defining thirty-six $6\times 6$ matrices of the form
\begin{equation}
(e_{IJ})_{KL}=\delta_{IK}\delta_{JL},\qquad I,J\ldots =1,\ldots 6\,
.
\end{equation}
Non-compact generators of $SO(3,3)$ are accordingly given by
\begin{equation}
Y_{ir}=e_{i,r+3}+e_{r+3,i},\qquad i,r=1,\ldots, 3\, .
\end{equation}
\indent We first truncate all of the nine scalars in $SO(3,3)/SO(3)\times SO(3)$ coset to scalars which are singlet under
$SO(3)_{\textrm{diag}}\subset SO(3)\times SO(3)$. There is only one singlet scalar corresponding to
the non-compact generator, see \cite{7D_flow} for more detail, 
\begin{equation}
Y_s=Y_{11}+Y_{22}+Y_{33}\, . 
\end{equation}
The coset representative can be written as
\begin{equation}
L=e^{\phi Y_s}\, .\label{L_SO3d}
\end{equation}
The scalar potential is straightforwardly computed to be
\begin{eqnarray}
V&=&\frac{1}{32}e^{-\sigma}\left[(g_1^2+g_2^2)\left[\cosh
(6\phi)-9\cosh(2\phi)\right]-8g_1g_2\sinh^3(2\phi)\phantom{e^{\frac{1}{2}}}\right.\nonumber
\\ & &\left.+8\left[g_2^2-g_1^2+64h^2e^{5\sigma}+32e^{\frac{5\sigma}{2}}h\left(g_1\cosh^3\phi-g_2\sinh^3\phi\right)\right]
\right].
\end{eqnarray}
\indent There are two supersymmetric $AdS_7$ critical points for this potential. 
\begin{itemize}
\item $AdS_7$ with $SO(4)$ symmetry:
\begin{equation}
\sigma =\phi=0,\qquad V_0=-240h^2\, .
\end{equation}
\item $AdS_7$ with $SO(3)_{\textrm{diag}}$ symmetry:
\begin{eqnarray}
\sigma &=&-\frac{1}{5}\ln
\left[\frac{g_2^2-256h^2}{g_2^2}\right],\qquad
\phi=\frac{1}{2}\ln\left[\frac{g_2-16h}{g_2+16h}\right],\nonumber
\\
V_0&=&-\frac{240g_2^{\frac{8}{5}}h^2}{(g_2^2-256h^2)^{\frac{4}{5}}}\,
.
\end{eqnarray}
\end{itemize}
We have set $g_1=-16h$ in order to make the dilaton $\sigma$ vanish at the $SO(4)$ critical point. This is equivalent to a redefinition of $\sigma$ by an appropriate shift. $V_0$ is the value of the scalar potential at the critical point. Holographic RG flow solutions interpolating between these two critical points and flows to non-conformal field theories have already been given in \cite{7D_flow}.

\subsection{Solutions flowing between $AdS_7$ vacua}
In this paper, we generalize the solutions studied in \cite{7D_flow} by including a non-vanishing three-form field in the solutions. Following \cite{7D_sol_Dibitetto}, we take the metric ansatz to be an $AdS_3\times S^3$-sliced domain wall
\begin{equation}
ds^2= e^{2U(r)}ds^2_{AdS_3}+e^{2V(r)}dr^2+e^{2W(r)}ds^2_{S^3}
\end{equation}
with the metrics on $AdS_3$ and $S^3$ given by
\begin{eqnarray}
ds^2_{AdS_3}&=&\frac{1}{\tau^2}\left[(dx^1)^2+\cosh^2x^1(dx^2)^2-(dt-\sinh x^1dx^2)^2\right],\nonumber \\
ds^2_{S^3}&=&\frac{1}{\kappa^2}\left[(d\theta_2)^2+\cos^2\theta_2(d\theta_3)^2+(d\theta_1+\sin \theta_2d\theta_3)^2\right].
\end{eqnarray} 
The seven-dimensional coordinates are taken to be $x^\mu=(x^a,r, x^m)$ with $a=0,1,2$ and $m=4,5,6$. We will also use $x^0=t$ and $x^3=r$ in the following analysis. The corresponding flat indices will be denoted by $\hat{\mu}=(\hat{a},\hat{3},\hat{m})$. The $S^3$ part is described by Hopf coordinates $x^m=(\theta_1,\theta_2,\theta_3)$. In the limit $\tau\rightarrow 0$ and $\kappa\rightarrow 0$, the $AdS_3$ and $S^3$ become flat Minkowski space and flat space $\mathbb{R}^3$, respectively.
\\
\indent With the vielbeins on $AdS_3$ and $S^3$ of the form
\begin{eqnarray}
e^{\hat{0}}&=& \frac{1}{\tau}(dt-\sinh{x^1}dx^2),\\
e^{\hat{1}}&=& \frac{1}{\tau}(\cos{t}dx^1-\sin{t}\cosh{x^1}dx^2),\\
e^{\hat{2}}&=& \frac{1}{\tau}(\sin{t}dx^1+\cos{t}\cosh{x^1}dx^2)
\end{eqnarray}
and
\begin{eqnarray}
e^{\hat{4}}&=& \frac{1}{\kappa}(d\theta_1+\sin{\theta_2}d\theta_3),\\
e^{\hat{5}}&=& \frac{1}{\kappa}(\cos{\theta_1}d\theta_2-\sin{\theta_1}\cos{\theta_2}d\theta_3),\\
e^{\hat{6}}&=& \frac{1}{\kappa}(\sin{\theta_1}d\theta_2+\cos{\theta_1}\cos{\theta_2}d\theta_3),
\end{eqnarray}
the spin connections take a simple form
\begin{eqnarray}
{{\omega_{\hat{a}}}^{\hat{a}}}_{\hat{3}}&=& e^{-V}U', \qquad \omega_{\hat{a}\hat{b}\hat{c}}=\frac{\tau}{2}e^{-U}\epsilon_{\hat{a}\hat{b}\hat{c}},\\
{{\omega_{\hat{m}}}^{\hat{m}}}_{\hat{3}}&=& e^{-V}W', \qquad  \omega_{\hat{m}\hat{n}\hat{p}}=\frac{\kappa}{2}e^{-W}\epsilon_{\hat{m}\hat{n}\hat{p}}\label{spin_connection2}
\end{eqnarray}
with $\epsilon_{\hat{0}\hat{1}\hat{2}}=\epsilon_{\hat{4}\hat{5}\hat{6}}=1$. We will use $'$ to denote the $r$-derivative throughout the paper. 
\\
\indent As in the usual domain wall solutions, the scalar fields $\sigma$ and $\phi$ are functions of only $r$ while the ansatz for the three-form field is taken to be 
\begin{equation}
C_{(3)}=k(r)\textrm{Vol}_{AdS_3}+l(r)\textrm{Vol}_{S^3}\label{3-form_ansatz}
\end{equation}
in which $\textrm{Vol}_{AdS_3}$ and $\textrm{Vol}_{S^3}$ are volume forms on $AdS_3$ and $S^3$, respectively. We will also set $A^I_{(1)}=0$ since, in this section, we are interested only in solutions with vanishing vector fields.
\\
\indent The ansatz for Killing spinors corresponding to the unbroken supersymmetry takes the form of
\begin{equation}
\epsilon=Y(r)\left[\cos{\theta(r)}\mathbf{1}_8+\sin{\theta(r)}\gamma^{\hat{0}\hat{1}\hat{2}}\right]\epsilon_0\label{Killing_spinor}
\end{equation}
with the constant spinor $\epsilon_0$ satisfying the projection condition
\begin{equation}
\gamma^{\hat{3}}\epsilon_0=\epsilon_0\, .\label{gamma_r_projection}
\end{equation} 
$Y(r)$ and $\theta(r)$ are functions of $r$ to be determined. 
\\
\indent To find supersymmetric solutions, we consider BPS equations obtained from supersymmetry transformations of fermionic fields $(\psi_\mu, \chi,\lambda^r)$. Using the Killing spinor \eqref{Killing_spinor} and the projection \eqref{gamma_r_projection}, we obtain two equations from $\delta\lambda^r=0$ conditions
\begin{eqnarray}
P^{ir}_{\hat{3}}\cos{2\theta}-\frac{1}{\sqrt{2}}e^{-\frac{\sigma}{2}}C^{ir}&=&0,\label{dlambda1_eq1}\\
P^{ir}_{\hat{3}}-\frac{1}{\sqrt{2}}e^{-\frac{\sigma}{2}}C^{ir}\cos{2\theta}&=&0\, .\label{dlambda1_eq2}
\end{eqnarray}
For the coset representative \eqref{L_SO3d}, we can readily compute $P^{ir}_\mu$ and $C^{ir}$. The result is given by
\begin{equation}
P^{ir}_{\hat{3}}=\phi'e^{-V}\delta^{ir}\qquad \textrm{and} \qquad C^{ir}=\sqrt{2}(g_1\cosh\phi-g_2\sinh\phi)\cosh{\phi}\sinh\phi\delta^{ir}\, .
\end{equation}
Note also that the three-form field does not enter the $\delta \lambda^r$ equations.
\\
\indent Compatibility between equations \eqref{dlambda1_eq1} and \eqref{dlambda1_eq2} implies $\cos(2\theta)=\pm 1$ leading to $\sin\theta=0$ or $\cos\theta=0$. Up to a redefinition of $\epsilon_0$ to $\tilde{\epsilon}_0=\gamma^{\hat{0}\hat{1}\hat{2}}\epsilon_0$ and a sign change in the projection condition \eqref{gamma_r_projection}, the two choices give equivalent BPS equations. For definiteness, we will choose $\sin\theta=0$ in the following analysis. This leads to the BPS equation for $\phi$
\begin{equation}
\phi'=e^{V-\frac{\sigma}{2}}(g_1\cosh\phi-g_2\sinh\phi)\cosh{\phi}\sinh\phi\, .\label{phi_eq1}
\end{equation}
The Killing spinor then takes a simpler form
\begin{equation}
\epsilon=Y(r)\epsilon_0\, .
\end{equation}
\indent We now consider $\delta \chi=0$ equation. This condition involves a contribution from the three-form field of the form $H_{\mu\nu\rho\sigma}\gamma^{\mu\nu\rho\sigma}\epsilon$. We will use the same convention for spinors and gamma matrices as in \cite{7D_sol_Dibitetto}. Using the relation $\gamma^{\hat{0}}\gamma^{\hat{1}}\gamma^{\hat{2}}\gamma^{\hat{3}}\gamma^{\hat{4}}\gamma^{\hat{5}}\gamma^{\hat{6}}=\mathbf{1}_8$ or more compactly $\epsilon_{\hat{a}\hat{b}\hat{c}}\gamma^{\hat{a}\hat{b}\hat{c}}\gamma^{\hat{r}}
=-\epsilon_{\hat{m}\hat{n}\hat{p}}\gamma^{\hat{m}\hat{n}\hat{p}}$, we find 
\begin{equation}
\frac{1}{4!}H_{\mu\nu\rho\sigma}\gamma^{\mu\nu\rho\sigma}\epsilon=(l'e^{-V-3W}-k'e^{-V-3U})\gamma^{\hat{0}\hat{1}\hat{2}}\epsilon\, .
\end{equation}    
Since there is no other term contributing $\gamma^{\hat{0}\hat{1}\hat{2}}$ matrix in the $\delta \chi$ variation, this term must vanish by itself. This can be achieved by setting
\begin{equation}
k'e^{-3U}=l'e^{-3W}\label{k_l_relation}
\end{equation}
which leads to the BPS equation for $\sigma$
\begin{equation}
\sigma'=-\frac{2}{5}e^{V-\frac{\sigma}{2}}\left[16he^{\frac{5}{2}\sigma}+g_1\cosh^3\phi-g_2\sinh^3\phi\right].\label{sigma_eq1}
\end{equation} 
\indent We then move on to the BPS equations from $\delta \psi_\mu$ conditions. After using the $\gamma^{\hat{r}}$ projection \eqref{gamma_r_projection} and the three-form ansatz \eqref{3-form_ansatz} in the conditions $\delta \psi_a=0$ and $\delta \psi_m=0$, we find two types of terms one with $\gamma^{\hat{0}\hat{1}\hat{2}}$ and the other with $\mathbf{1}_8$. The former gives rise to the BPS equations for $k$ and $l$ 
\begin{equation}
k'=\frac{\tau}{\sqrt{2}}e^{2U+\sigma+V},\qquad l'=\frac{\kappa}{\sqrt{2}}e^{2W+\sigma+V}\label{k_l_eq}
\end{equation}
while the latter gives the corresponding equations for $U$ and $W$
\begin{equation}
U'=W'=\frac{1}{5}e^{V-\frac{\sigma}{2}}\left[4he^{\frac{5}{2}\sigma}-g_1\cosh^3\phi+g_2\sinh^3\phi\right].
\end{equation}
The last equation implies that $U=W+C$ for a constant $C$. In order to find solutions interpolating between $AdS_7$ vacua, we require that the solutions be asymptotically locally $AdS_7$ at which $U=W$. This implies that $C=0$ or $U=W$. 
\\
\indent Using this relation in equation \eqref{k_l_relation}, we find that $k'=l'$ or $k=l+\tilde{C}$ for some constant $\tilde{C}$. This constant can be set to zero by a suitable redefinition of $k$ and $l$. We will accordingly set $k=l$. With all these, equation \eqref{k_l_eq} gives
\begin{equation} 
\tau=\kappa\, .
\end{equation}
In summary, we end up with the BPS equations for the warped factor $U$ and $k$ in the form of
\begin{eqnarray}
U'&=&\frac{1}{5}e^{V-\frac{\sigma}{2}}\left[4he^{\frac{5}{2}\sigma}-g_1\cosh^3\phi+g_2\sinh^3\phi\right],\label{U_eq1}\\
k'&=&\frac{\kappa}{\sqrt{2}}e^{2U+\sigma+V}\, .\label{k_eq1}
\end{eqnarray} 
\indent It should be noted that the contribution from $C_{(3)}$ is cancelled by the spin connections on $AdS_3$ and $S^3$. Therefore, for non-vanishing $C_{(3)}$ and $k=l$, there can be no background with $\textrm{Mink}_3$ and $\mathbb{R}^3$. This is perfectly in agreement with a similar solution considered in \cite{7D_sol_Dibitetto} but without the scalar from vector multiplets. It can also be easily checked that any solutions to the above BPS equations solve all the field equations.
\\
\indent We finally consider the equation from $\delta \psi_3$ condition. This gives the BPS equation for $Y(r)$ 
\begin{equation}
Y'=\frac{1}{2}YU'
\end{equation}
which can be solved by a solution $Y\sim e^{\frac{U}{2}}$.
\\
\indent We are now in a position to solve all of the BPS equations. To find an analytic solution, we first choose a function $V(r)=\frac{\sigma}{2}$. This is equivalent to changing to a new radial coordinate $\tilde{r}$ defined by the relation $\frac{d\tilde{r}}{dr}=e^{-\frac{\sigma}{2}}$ in \cite{7D_flow}. The procedure is very similar to that used in \cite{7D_flow}, so we will not repeat all the details here. After choosing $V(r)=\frac{\sigma}{2}$, we obtain the solution for \eqref{phi_eq1}
\begin{equation}
g_1g_2r=g_2\ln (1-e^{2\phi})-g_1\ln (1+e^{2\phi})+\frac{(g_1+g_2)^2}{g_1-g_2}\ln[ g_1+g_2+(g_1-g_2)e^{2\phi}]
\end{equation}
where an irrelevant additive integration constant has been neglected.
\\  
\indent By treating $U$, $\sigma$ and $k$ as functions of $\phi$, we find the solution of equations \eqref{sigma_eq1}, \eqref{U_eq1} and \eqref{k_eq1}    
\begin{eqnarray}
\sigma&=&\frac{2}{5}\ln\left[\frac{g_1g_2}{16h(g_1\sinh\phi-g_2\cosh\phi)}\right],\\
U&=&\frac{1}{4}\phi-\frac{1}{8}\sigma-\frac{1}{4}\ln (e^{4\phi}-1)+\frac{1}{4}\ln[g_1+g_2+(g_1-g_2)e^{2\phi}],\\
k&=&\frac{\tau}{4}\left[\frac{g_1}{g_2}+\frac{g_2}{g_1}-2\coth(2\phi)\right]
\end{eqnarray}
in which irrelevant integration constants in $U$ and $k$ have been removed. The integration constant in $\sigma$ is however important and has been chosen such that the solution for $\sigma$ interpolates between the two supersymmetric $AdS_7$ critical points, see \cite{7D_flow} for more detail.   
\\
\indent As $r\rightarrow \pm \infty$, the solution is asymptotic to the $AdS_7$ critical points with
\begin{equation}
U\sim 4hr,\qquad \sigma\sim\phi\sim 0,\qquad F_{\hat{0}\hat{1}\hat{2}\hat{r}}\sim F_{\hat{r}\hat{4}\hat{5}\hat{6}}\sim 0
\end{equation}
for $r\rightarrow \infty$, and
\begin{eqnarray}
& &U\sim 4h\left(\frac{g_2^2}{g_2^2-256h^2}\right)^{\frac{2}{5}}r,\qquad  \sigma\sim \frac{1}{5}\ln\left[\frac{g^2_2}{g^2_2+256h^2}\right],\nonumber \\
& & \phi\sim \frac{1}{2}\ln \left[\frac{g_2-16h}{g_2+16h}\right],\qquad F_{\hat{0}\hat{1}\hat{2}\hat{r}}\sim F_{\hat{r}\hat{4}\hat{5}\hat{6}}\sim 0
\end{eqnarray}
for $r\rightarrow -\infty$. In these equations, we have set $g_1=-16h$. 
\\
\indent It should be noted that the four-form field strength does not actually vanish in the limit $r\rightarrow \pm\infty$ as can be seen from the BPS equation for $k'$. Moreover, the existence of $C_{(3)}$ is needed to support the $AdS_3$ and $S^3$ factors as mentioned above. However, its effect in the limit $r\rightarrow \pm \infty$ is highly suppressed compared to the scalar potential. The solution is then asymptotically locally $AdS_7$ as $r\rightarrow \pm\infty$.  

\subsection{Solutions with known higher dimensional origin}
For a particular case of $g_2=g_1$, solutions of the $N=2$ gauged supergravity can be uplifted to eleven dimensions. The corresponding reduction ansatz has been constructed in \cite{7D_from_11D}. Setting $g_2=g_1$, we obtain the BPS equations  
\begin{eqnarray}
\phi'&=&e^{V-\frac{\sigma}{2}-\phi}g_1\cosh\phi \sinh\phi,\\
\sigma'&=&-\frac{2}{5}e^{V-\frac{\sigma}{2}}\left[16he^{\frac{5}{2}\sigma}+g_1\cosh^3\phi-g_1\sinh^3\phi\right],\\
U'&=&\frac{1}{5}e^{V-\frac{\sigma}{2}}\left[4he^{\frac{5}{2}\sigma}-g_1\cosh^3\phi+g_1\sinh^3\phi\right],\\
k'&=&\frac{\kappa}{\sqrt{2}}e^{2U+\sigma+V}\, .
\end{eqnarray}
It can be clearly seen from the $\phi'$ equation that there is only one supersymmetric $AdS_7$ background at $\phi=0$. The solutions interpolating between this $AdS_7$ and physically acceptable, singular geometries dual to non-conformal field theories in the case of $k=0$ have already been studied in \cite{7D_from_11D}. In this paper, we will give the solution with non-vanishing three-form field. This solution can be found by the analysis similar to the previous case. The resulting solution is given by 
\begin{eqnarray}
g_1r&=&2\tan^{-1}e^\phi-2\tanh^{-1}e^\phi,\\
\sigma&=&\frac{2}{5}\phi-\frac{2}{5}\ln \left[1-12C_1(e^{4\phi}-1)\right],\\
U&=&\frac{1}{5}\phi-\frac{1}{4}\ln(e^{4\phi}-1)+\frac{1}{20}\ln [1-12C_1(e^{4\phi}-1)],\\
k&=&\frac{\tau}{2h}\left(\frac{h^4}{2^9g_1^4}\right)^{\frac{1}{10}}\sqrt{\frac{1-12C_1(e^{4\phi}-1)}{e^{4\phi}-1}}\, .
\end{eqnarray}
It can be seen that $\phi$ diverges at a finite value of $r$. Therefore, the solution is singular at this point. Without loss of generality, we can shift the coordinate $r$ such that the singularity occurs at $r=0$. The integration constant $C_1$ controls the behavior near the singularity, see \cite{7D_from_11D} for more detail.
\\
\indent For $C_1=0$, the solution near $r=0$ becomes
\begin{eqnarray}
& &\phi\sim -\ln (4hr),\qquad \sigma\sim -\frac{2}{5}\ln (4hr),\qquad k\sim e^{-2\phi}\sim (4hr)^2,\nonumber \\
& &ds^2_7=(4hr)^2\left(ds^2_{AdS_3}+ds^2_{S^3}\right)+(4hr)^{-\frac{1}{5}}dr^2
\end{eqnarray}
in which we have set $g_1=-16h$. For $C_1\neq 0$, we find
\begin{eqnarray}
& &\phi\sim -\ln (4hr),\qquad \sigma\sim \frac{6}{5}\ln (4hr),\qquad k\sim \textrm{constant},\nonumber \\
& &ds^2_7=(4hr)^{\frac{3}{4}}\left(ds^2_{AdS_3}+ds^2_{S^3}\right)+(4hr)^{\frac{3}{5}}dr^2\, .
\end{eqnarray}
As pointed out in \cite{7D_from_11D}, all of these singularities are physically acceptable since the scalar potential is bounded from above, in this case $V\rightarrow -\infty$, as required by the criterion in \cite{Gubser_singularity}.
\\
\indent In this case, the solution can be embedded in eleven dimensions by using the reduction ansatz in \cite{7D_from_11D}. For convenience, we give a brief review of this result in appendix \ref{reduction_ansatz}. The nine scalars from vector multiplets can be equivalently described by $SL(4,\mathbb{R})/SO(4)$ coset due to the isomorphism $SL(4,\mathbb{R})\sim SO(3,3)$. For the $SO(3)_{\textrm{diag}}$ singlet scalar, we find the $SL(4,\mathbb{R})/SO(4)$ coset representative 
\begin{equation}
{\mc{V}_\alpha}^R=\textrm{diag}(e^{\frac{\phi}{2}},e^{\frac{\phi}{2}},e^{\frac{\phi}{2}},e^{-\frac{3\phi}{2}})
\end{equation}
which gives a symmetric $4\times 4$ matrix with unit determinant  
\begin{equation}
\tilde{T}_{\alpha\beta}=\textrm{diag}(e^\phi,e^\phi,e^\phi,e^{-3\phi})=(\delta_{ab}e^\phi,e^{-3\phi}).
\end{equation}
In the remaining parts of this section, we will use indices $a,b=1,2,3$ to denote coordinates $\hat{\mu}^a$ on the internal $S^2$ with $\hat{\mu}^a\hat{\mu}^a=1$. We will also use the $S^3$ coordinates $\mu^\alpha=(\cos\psi \hat{\mu}^a,\sin\psi)$ satisfying $\mu^\alpha\mu^\alpha=1$. 
\\
\indent With all these and the seven-dimensional fields given previously, we obtain the eleven-dimensional metric
\begin{eqnarray}
d\hat{s}^2_{11}&=&\Delta^{\frac{1}{3}}\left[e^{2U}\left(ds^2_{AdS_3}+ds^2_{S^3}\right)+e^{2V}dr^2\right]\nonumber \\
& &+\frac{1}{32h^2}\Delta^{-\frac{2}{3}}e^{-2\sigma}\left[\cos^2\xi+e^{\frac{5}{2}\sigma}\sin^2\xi (e^{-\phi}\cos^2\psi+e^{3\phi}\sin^2\psi)\right]d\xi^2\nonumber \\
& &+\frac{1}{64h^2}\Delta^{-\frac{2}{3}}e^{\frac{\sigma}{2}}\sin\xi\sin\psi\cos\psi(e^{-\phi}-e^{3\phi})d\xi d\psi \nonumber \\
& &+\frac{1}{128h^2}\Delta^{-\frac{2}{3}}e^{\frac{\sigma}{2}}\cos^2\xi \left[(e^{3\phi}\cos^2\psi+e^{-\phi}\sin^2\xi)d\psi^2+e^{-\phi}\cos^2\psi d\Omega^2_2\right]\nonumber \\
& &
\end{eqnarray}
with the warped factor given by
\begin{equation}
\Delta=e^{-\frac{\sigma}{2}}\cos^2\xi (e^\phi\cos^2\psi+e^{-3\phi}\sin^2\psi)+e^{2\sigma}\sin^2\xi,
\end{equation}
and the metric on a unit two-sphere can be written as $d\Omega^2_2=d\hat{\mu}^a d\hat{\mu}^a$. It should be noted that the $S^2$ in the internal $S^3$ is unchanged. Its isometry corresponds to the unbroken $SO(3)_{\textrm{diag}}$ symmetry.
\\
\indent The four-form field strength of eleven-dimensional supergravity is given by
\begin{eqnarray}
\hat{F}_{(4)}&=&\sin\xi \sqrt{2}k'(dr\wedge \textrm{Vol}_{AdS_3}+dr\wedge \textrm{Vol}_{S^3})\nonumber \\
& &-\frac{\sqrt{2}}{8h}e^{-2\sigma}\cos\xi k'e^{-V}(\textrm{Vol}_{AdS_3}+\textrm{Vol}_{S^3})\wedge d\xi\nonumber\\
& &+\frac{1}{(8h)^3}\Delta^{-2}U\cos^3\xi \cos^2\psi d\xi\wedge d\psi\wedge \epsilon_{(2)}\nonumber \\
& &+\frac{1}{(8h)^3}\Delta^{-2}e^{\frac{3}{2}\sigma}\sin\xi\cos^4\xi\cos^2\psi\left[e^\phi\cos^2\psi\left(\phi'-\frac{5}{2}\sigma'\right)\right. \nonumber \\
& &\left.-e^{-3\phi}\sin^2\psi \left(\frac{5}{2}\sigma'+3\phi'\right)\right]dr\wedge d\psi\wedge \epsilon_{(2)}\nonumber \\
& &-\frac{1}{(8h)^3}\Delta^{-2}\cos^2\xi\cos^3\psi\sin\psi\left[\left[4e^{-2\phi-\sigma}\cos^3\xi +e^{\frac{3}{2}\sigma}(e^\phi+3e^{-3\phi})\right]\phi'\phantom{\frac{5}{2}}\right.\nonumber \\
& &\left.-\frac{5}{2}\sin^2\xi e^{\frac{3}{2}\sigma}(e^\phi-e^{-3\phi})\sigma'\right]dr\wedge d\xi\wedge \epsilon_{(2)}
\end{eqnarray}
where $\epsilon_{(2)}=\frac{1}{2}\epsilon_{abc}\hat{\mu}^ad\hat{\mu}^bd\hat{\mu}^c$ is the volume form on $S^2$. In this equation, we have also used $\epsilon_{abc4}=\epsilon_{abc}$. The scalar function $U$ is given by
\begin{eqnarray}
U&=&\sin^2\xi (e^{4\sigma}-3e^{\phi+\frac{3}{2}\sigma}-e^{\frac{3}{2}\sigma-3\phi})-\cos^2\xi\left[e^{\frac{3}{2}\sigma}(e^\phi\cos^2\psi+e^{-3\phi}\sin^2\psi) \right.\nonumber \\
& & \left.\phantom{e^{\frac{3}{2}\sigma}} +e^{-\sigma}(e^{2\phi}\cos^2\psi+3e^{2\phi}\sin^2\psi+e^{-2\phi}\cos^2\psi-e^{-6\phi}\sin^2\psi)\right].
\end{eqnarray}
\indent Similar to the discussion in \cite{7D_sol_Dibitetto}, we expect the uplifted solution to describe eleven-dimensional configurations involving M$2$-M$5$-brane bound states due to the dyonic profile of $C_{(3)}$. It is also interesting to consider the $(00)$-component of the eleven-dimensional metric
\begin{equation}
\hat{g}_{00}=-\frac{1}{\kappa^2}\Delta^{\frac{1}{3}}e^{2U(r)}\, .
\end{equation} 
Near the singularity at $r= 0$, we find that 
\begin{equation} 
\hat{g}_{00}\sim (4hr)^{\frac{26}{15}}\rightarrow 0\qquad
\textrm{and}\qquad  \hat{g}_{00}\sim (4hr)^{\frac{13}{60}}\rightarrow 0
\end{equation}
for $C_1=0$ and $C_1\neq 0$, respectively. According to the criterion of \cite{maldacena_nogo}, the singularities are physical in agreement with the seven-dimensional results obtained from the criterion of \cite{Gubser_singularity}. We then expect that the solution holographically describes a two-dimensional conformal defect in six-dimensional $N=(1,0)$ SCFT with known M-theory origin.  

\section{Domain walls with the three-form and vector fields}\label{DW_3form_vector}
In this section, we consider more general solutions with non-vanishing vector fields. We first choose an appropriate ansatz for the $SO(4)\sim SO(3)\times SO(3)$ gauge fields. As in \cite{7D_sol_Dibitetto}, we will take this ansatz in the form of  
\begin{equation}
A^I_{(1)}=A^I_id\theta_i
\end{equation}
in which the components $A^I_i$ will be functions of only the radial coordinate $r$. Explicitly, these components are given by
\begin{eqnarray}
A^{i}_j&=&-\frac{e^{-W}\kappa}{2}A(r)\delta^i_j,\\
A^r_i&=&-\frac{e^{-W}\kappa}{2}B(r)\delta^r_i\, .
\end{eqnarray}
It is now straightforward to compute the field strength tensors $F^i_{(2)}={L_I}^iF^I_{(2)}$ and $F^r_{(2)}={L_I}^rF^I_{(2)}$. Non-vanishing components of these tensors are given by
\begin{eqnarray}
F^i_{3j}&=&\mathbf{f}\delta^i_j,\qquad F^i_{jk}=\mathbf{g}\epsilon_{ijk},\\
F^r_{3i}&=&\bar{\mathbf{f}}\delta^r_i,\qquad F^r_{jk}=\bar{\mathbf{g}}\delta^r_i\epsilon_{ijk}
\end{eqnarray}
where
\begin{eqnarray}
\mathbf{f}&=&-e^{-V-W}\frac{\kappa}{2}\left[A'\cosh{\phi}+B'\sinh{\phi}\right],\\
\bar{\mathbf{f}}&=&-e^{-V-W}\frac{\kappa}{2}\left[A'\sinh{\phi}+B'\cosh{\phi}\right],\\
\mathbf{g}&=&-e^{-2W}\frac{\kappa^2}{4}\left[A(2-g_1A)\cosh{\phi}+B(2+g_2B)\sinh{\phi}\right],\\
\bar{\mathbf{g}}&=&-e^{-2W}\frac{\kappa^2}{4}\left[A(2-g_1A)\sinh{\phi}+B(2+g_2B)\cosh{\phi}\right].
\end{eqnarray}
To implement $SO(3)_{\textrm{diag}}$, we will set 
\begin{equation}
g_2B=-g_1A\, .\label{SO3D_con}
\end{equation}
\indent We still use the ansatz for the Killing spinor as given in \eqref{Killing_spinor} and the projection \eqref{gamma_r_projection}. Due to the extra contributions from non-vanishing gauge fields, we need more projectors
\begin{equation}
\gamma^{\hat{5}\hat{6}}\epsilon=-i\sigma^1\epsilon,\qquad \gamma^{\hat{4}\hat{6}}\epsilon=i\sigma^2\epsilon,\qquad \gamma^{\hat{4}\hat{5}}\epsilon=-i\sigma^3\epsilon,
\end{equation}
The second condition is just the Symplectic-Majorana condition. Therefore, the BPS solutions (if exist) will preserve only two supercharges or $\frac{1}{8}$-BPS after imposing the projection \eqref{gamma_r_projection}. 
\\
\indent With all these, we can now set up the BPS equations. By the relation \eqref{SO3D_con}, the composite connection along $S^3$ takes a very simple form
\begin{equation}
Q_{ijk}=\omega_{i+3,j+3,k+3}
\end{equation}
where $\omega_{i+3,j+3,k+3}$ is the spin connection given in \eqref{spin_connection2}. Using the same procedure as in the previous section, we find the following set of BPS equations
\begin{eqnarray}
U'&=&\frac{e^V}{60\cos{2\theta}}\left[2\left[12he^{2\sigma}+\frac{e^{-\frac{\sigma}{2}}}{\sqrt{2}}C\right](3\cos{4\theta}-1)+18e^{\frac{\sigma}{2}}\mathbf{g}(\cos{4\theta}-3)\right. \nonumber \\
& &\left.\phantom{\frac{e^{-\frac{\sigma}{2}}}{\sqrt{2}}} +24e^{-U}\tau\sin{2\theta}+18e^{-W}\kappa(g_1A-1)\sin{4\theta}\right],\label{U_eq2}\\
W'&=& -\frac{e^V}{30\cos{2\theta}}\left[2\left[12he^{2\sigma}+\frac{e^{-\frac{\sigma}{2}}}{\sqrt{2}}C\right](\cos{4\theta}-2)+6e^{\frac{\sigma}{2}}\mathbf{g}(\cos{4\theta}-8)\right. \nonumber \\
& &\left.\phantom{\frac{e^{-\frac{\sigma}{2}}}{\sqrt{2}}}+18e^{-U}\tau\sin{2\theta}+6e^{-W}\kappa(g_1A-1)\sin{4\theta}\right],\\
Y'&=& \frac{e^VY}{120\cos{2\theta}}\left[2\left[12he^{2\sigma}+\frac{e^{-\frac{\sigma}{2}}}{\sqrt{2}}C\right](3\cos{4\theta}-1)+18e^{\frac{\sigma}{2}}\mathbf{g}(\cos{4\theta}-3)\right.\nonumber \\
& &\left.\phantom{\frac{e^{-\frac{\sigma}{2}}}{\sqrt{2}}} +24e^{-U}\tau\sin{2\theta}+18e^{-W}\kappa(g_1A-1)\sin{4\theta}\right],\\
\theta'&=& \frac{e^V}{4}\left[-2\left[4he^{2\sigma}+\frac{e^{-\frac{\sigma}{2}}}{\sqrt{2}}C\right]\sin{2\theta}-6e^{\frac{\sigma}{2}}\mathbf{g}\sin{2\theta}+6e^{-U}\tau\right. \nonumber  \\
& &\left.\phantom{\frac{e^{-\frac{\sigma}{2}}}{\sqrt{2}}} +6e^{-W}\kappa(g_1A-1)\cos{2\theta}\right],\label{theta_eq}\\
k'&=& \frac{e^{3U+V}e^\sigma}{3\sqrt{2}}\left[2\left[12he^{2\sigma}+\frac{e^{-\frac{\sigma}{2}}}{\sqrt{2}}C\right]\tan{2\theta}+18e^{\frac{\sigma}{2}}\mathbf{g}\tan{2\theta}\right. \nonumber 
\\
& &\left.\phantom{\frac{e^{-\frac{\sigma}{2}}}{\sqrt{2}}}-6e^{-U}\tau\sec{2\theta}-9e^{-W}\kappa(g_1A-1)\right],\label{k_eq}
\end{eqnarray}
\begin{eqnarray}
l'&=&\frac{1}{\sqrt{2}}e^{3W+V}e^\sigma\left[e^{-U}\tau-8he^{2\sigma}\sin{2\theta}\right],\label{l_eq}\\
\sigma'&=& -\frac{e^V}{30\cos{2\theta}}\left[48he^{2\sigma}(\cos{4\theta}+3)-\sqrt{2}Ce^{-\frac{\sigma}{2}}(3\cos{4\theta}-1)-24e^{-U}\tau\sin{2\theta}\right.\nonumber \\
& &\left.\phantom{\sqrt{2}}-18e^{\frac{\sigma}{2}}\mathbf{g}(\cos{4\theta}-3)-18e^{-W}\kappa(g_1A-1)\sin{4\theta}\right],\\
\phi'&=& e^V\left[\frac{e^{-\frac{\sigma}{2}}}{\sqrt{2}}\mathcal{C}-e^{\frac{\sigma}{2}}\bar{\mathbf{g}}\right]\cos{2\theta},\label{phi_eq}\\
A'&=&-\frac{2g_2e^{V+W-\frac{\sigma}{2}}}{\kappa (g_1\sinh\phi-g_2\cosh\phi)}\left[\frac{1}{3\sqrt{2}}Ce^{-\frac{\sigma}{2}}\sin{2\theta}+e^{\frac{\sigma}{2}}\mathbf{g}\sin{2\theta}\right. \nonumber \\
& &\left.\phantom{\frac{1}{3\sqrt{2}}}-\left[e^{-U}\tau+e^{-W}\kappa(g_1A-1)\cos{2\theta}\right]\right]\label{A_eq2}
\end{eqnarray}
where the quantities $C$ and $\mc{C}$ are defined by
\begin{eqnarray}
C&=&-3\sqrt{2}\left(g_1\cosh^3{\phi}-g_2\sinh^3{\phi}\right),\\
\mathcal{C}&=&\frac{1}{\sqrt{2}}\sinh(2\phi)\left(g_1\cosh{\phi}-g_2\sinh{\phi}\right).
\end{eqnarray}
\indent In addition to these flow equations, there is an algebraic constraint arising from the fact that the supersymmetry transformations from the gravity multiplet ($\delta \psi_\mu$ and $\delta \chi$) and those from the vector multiplets ($\delta \lambda^r$) lead to different BPS equations for $A$. Consistency between these two equations results in a constraint
\begin{eqnarray}
0&=& e^{\frac{\sigma}{2}}\sin{2\theta}\left[\left(\frac{e^{-\sigma}}{3\sqrt{2}}C+\mathbf{g}\right) +\frac{g_1\sinh{\phi}-g_2\cosh{\phi}}{g_1\cosh{\phi}-g_2\sinh{\phi}}\left(-\frac{e^{-\sigma}}{\sqrt{2}}\mathcal{C}+\bar{\mathbf{g}}\right)\right]
\nonumber \\
& &+e^{-W}\kappa(1-g_1A)\cos{2\theta}-e^{-U}\tau\, .\label{constraint_SUSY}
\end{eqnarray} 
This means supersymmetric solutions must satisfy the above BPS equations as well as the constraint \eqref{constraint_SUSY} in order for the Killing spinors to exist. We have explicitly verified that the BPS equations \eqref{U_eq2} to \eqref{A_eq2} together with the constraint \eqref{constraint_SUSY} imply all of the field equations.
\\
\indent However, it turns out that there are no solutions with non-trivial three-form and vector fields satisfying the BPS equations \eqref{U_eq2} to \eqref{A_eq2} and the constraint \eqref{constraint_SUSY}. This can be readily checked by differentiating equation \eqref{constraint_SUSY} and substituting the BPS equations \eqref{U_eq2} to \eqref{A_eq2}. The result gives the following condition
\begin{equation}
0=e^{-2U-\frac{\sigma}{2}}k'\mathbf{g}-\frac{2e^{U+\frac{\sigma}{2}}(g_2^2-g_1^2)\phi'\mathbf{f}}{(g_2\sinh\phi-g_1\cosh\phi)(g_2\cosh\phi-g_1\sinh\phi)}\label{D_constraint}
\end{equation}
which can not be satisfied by the BPS equations \eqref{k_eq} and \eqref{phi_eq}. This implies that $\phi$ and $k$ cannot flow independently but relate to each other by equation \eqref{D_constraint}. 
\\
\indent Note that relation \eqref{D_constraint} is trivially satisfied for $A=0$ which gives $\mathbf{f}=\mathbf{g}=0$. This results in the BPS equations considered in the previous section. Another possibility is to set $k'=0$ and $g^2_2=g^2_1$. However, with $k'=0$ and the constraint \eqref{constraint_SUSY}, equation \eqref{k_eq} implies $\theta=0$ and $A=\frac{1}{g_1}$ for generic values of $\sigma$ and $\phi$. Although this result is compatible with the BPS equations given in \eqref{theta_eq} and \eqref{A_eq2}, it also leads to $l'=0$ as can be seen from equation \eqref{l_eq} after using the constraint \eqref{constraint_SUSY}. The three-form field then has vanishing field strength. In this case, the gauge fields do not depend on the radial coordinate $r$. This type of solutions has already been studied in \cite{7D_twist} in the context of twisted compactifications. Therefore, we conclude that there are no supersymmetric solutions with non-vanishing $SO(3)_{\textrm{diag}}$ gauge fields and non-trivial three-form field.   

\section{Conclusions and discussions}\label{conclusion}
We have studied supersymmetric solutions of matter-coupled $N=2$ gauged supergravity in seven dimensions with $SO(4)\sim SO(3)\times SO(3)$ gauge group. The resulting solutions are generalizations of the previously known solutions of $N=2$ gauged supergravity in the sense that all possible bosonic fields, from both gravity and vector multiplets, are considered. These solutions take the form of asymptotically locally $AdS_7$ solutions and should be useful in the holographic study of $N=(1,0)$ six-dimensional SCFTs. 
\\
\indent For vanishing vector fields, we have found analytic solutions to the BPS equations for all the fields that are singlets of the residual $SO(3)_{\textrm{diag}}\subset SO(3)\times SO(3)$ symmetry. For special values of $SO(3)\times SO(3)$ gauge couplings, namely $g_2=g_1$, the solutions can be uplifted to eleven dimensions. We have performed this uplift and given the explicit form of the eleven-dimensional metric and the four-form field strength tensor. Unlike the solutions found in \cite{7D_sol_Dibitetto}, we have found that the solutions to the matter-coupled gauged supergravity are more restrictive. As a result, only solutions in the form of $AdS_3\times S^3$ sliced-domain walls are possible. From a general notion of the AdS/CFT correspondence and the recent result in \cite{6D_surface_Dibitetto}, we expect these solutions to describe some supersymmetric, two-dimensional, conformal defects in $N=(1,0)$ SCFT with flavor symmetry $SO(3)$. It is interesting to find the dual descriptions of these solutions in the $N=(1,0)$ SCFT. A generalization of these solutions to include more scalars, such as those invariant under smaller residual symmetries, is straightforward since the three-form field does not couple directly to scalars from vector multiplets. 
\\
\indent We have also derived a set of first-order flow equations together with an algebraic constraint for the case of non-vanishing $SO(3)_{\textrm{diag}}$ gauge fields. In this case, we have performed the analysis and argued that supersymmetric solutions do not exist at least within the truncation considered here. This is due to the fact that, in general, the constraint, arising from the supersymmetry variations of the gaugini, is violated by the solutions of the flow equations. 
\\
\indent Solutions in $N=2$ gauged supergravity with other gauge groups and the slicing different from $AdS_3\times S^3$ are worth considering. Similar solutions in the maximal $N=4$ gauged supergravity in seven dimensions are also of particular interest. These would describe lower dimensional defects within $N=(2,0)$ SCFT dual to the $AdS_7\times S^4$ solution of M-theory. We hope to come back to these issues in future works.   
\vspace{0.5cm}\\
{\large{\textbf{Acknowledgement}}} \\
We would like to thank Giuseppe Dibitetto and Davide Cassani for helpful correspondences and clarifications. This work is supported by The Thailand Research Fund (TRF) under grant RSA5980037.
\appendix
\section{Bosonic field equations of $N=2$ gauged supergravity coupled to vector multiplets}\label{field_equations}
In this appendix, we give all of the bosonic field equations obtained from the Lagrangian given in \eqref{7D_action}. These equations read
\begin{eqnarray}
d\left(e^{-2\sigma}*H_{(4)}\right)+8h H_{(4)}-\frac{1}{\sqrt{2}}F^I_{(2)}\wedge F^I_{(2)}&=&0,\label{7D4form_eq}\\
\frac{5}{4}d*d\sigma-\frac{1}{2}e^{\sigma}a_{IJ}*F^I_{(2)}\wedge F^J_{(2)}+e^{-2\sigma}*H_{(4)}\wedge H_{(4)}\qquad & &\nonumber \\
+\left[\frac{1}{4}e^{-\sigma}\left(C^{ir}C_{ir}-\frac{1}{9}C^2\right)+2\sqrt{2}he^{\frac{3}{2}\sigma}C-64 h^2 e^{4\sigma}\right]\epsilon_{(7)}&=&0,\label{7Ddilato_eq}\\
D(e^\sigma a_{IJ}*F^I_{(2)})-\sqrt{2}H_{(4)}\wedge F^J_{(2)}+*P^{ir}f_{IJ}^{\phantom{sas}K}L^I_{\phantom{s}r}L_{iK}&=&0,\label{7DYM_eq}\\
D*P^{ir}-2e^\sigma
L^i_{\phantom{s}I}L^r_{\phantom{s}J}*F^I_{(2)}\wedge
F^J_{(2)}\qquad\qquad
& &\nonumber \\
-\left[\frac{1}{\sqrt{2}}e^{-\sigma}C_{jr}C^{krs}\epsilon^{ijk}+4\sqrt{2}he^{\frac{3\sigma}{2}}C_{ir}\right]\epsilon_{(7)}&=&0,\label{7Dscalar_eq}\\
R_{\mu\nu}-\frac{5}{4}\pd_\mu \sigma \pd^\mu \sigma -P^{ir}_\mu P_{\nu ir}-\frac{2}{5}g_{\mu\nu}V\qquad \qquad  & &\nonumber \\ 
-a_{IJ}e^\sigma \left(F^I_{\mu\rho}{F^{J}_\nu}^\rho-\frac{1}{10}g_{\mu\nu}F^I_{\rho\sigma}F^{J\rho\sigma}\right)\qquad & &\nonumber \\
-\frac{1}{6}e^{-2\sigma}\left(H_{\mu\rho\sigma\lambda}{H_\nu}^{\rho\sigma\lambda}-\frac{3}{20}g_{\mu\nu}H_{\rho\sigma\lambda\tau}H^{\rho\sigma\lambda\tau}\right)&=&0\label{7DEinstein_eq}
\end{eqnarray}
where $C_{irs}$ is defined by
\begin{equation}
C_{irs}=f_{IJ}^{\phantom{sad}K}L^I_{\phantom{s}r}L^J_{\phantom{s}s}L_{Ki}\, .
\end{equation}
\indent The Yang-Mills equation \eqref{7DYM_eq} can also be written in terms of
$C^{ir}$ and $C^{irs}$ by using the relation
\begin{equation}
f_{IJ}^{\phantom{sad}K}L^I_{\phantom{s}r}L_{iK}=-\frac{1}{2\sqrt{2}}\epsilon^{ijk}C^{jr}L^k_{\phantom{s}J}-
C^{irs}L_{sJ}\, .
\end{equation}
In deriving the scalar field equation \eqref{7Dscalar_eq}, it is useful to adopt the following projections, given in \cite{Eric_N2_7D}, in the form of  
\begin{eqnarray}
\delta L^i_{\phantom{s}I}&=&X^i_{\phantom{s}r}L^r_{\phantom{s}I}+X^i_{\phantom{s}j}L^j_{\phantom{s}I},\nonumber \\
\delta L^r_{\phantom{s}I}&=&{X^r}_s{L^s}_I+{X^r}_i{L^i}_I\, .
\end{eqnarray}
With these relations, variations with respect to the scalar fields lead to the following results
\begin{eqnarray}
\delta C^2&=&-6CC^{ir}X_{ir},\\
\delta
(C^{ir}C_{ir})&=&2\sqrt{2}C_{js}C^{krs}\epsilon^{ijk}{X^i}_r-\frac{2}{3}C_{ir}CX^i_{\phantom{i}r}\,
.
\end{eqnarray} 

\section{Reduction ansatz from eleven dimensions}\label{reduction_ansatz}
In this appendix, we summarize all relevant formulae for embedding seven-dimensional solutions in eleven-dimensional supergravity. The reduction ansatz given in \cite{7D_from_11D} is obtained from a truncation of the $S^4$ reduction giving rise to the maximal $N=4$ $SO(5)$ gauged supergravity \cite{11D_to_7D_Nastase}. The reduction gives an effective seven-dimensional $N=2$ gauged supergravity coupled to three vector multiplets and $SO(4)$ gauge group. Due to the isomorphism $SO(3,3)\sim SL(4,\mathbb{R})$, the nine scalars from the vector multiplets can be equivalently parametrized by $SL(4,\mathbb{R})/SO(4)$ coset manifold. We follow all the conventions of \cite{7D_from_11D} to which the reader is referred for more detail. 
\\
\indent Let ${\mc{V}_\alpha}^R$ be $SL(4,\mathbb{R})/SO(4)$ coset representative with $\alpha, R=1,2,3,4$. The bosonic part of the $N=2$ gauged supergravity is more conveniently described by a symmetric $4\times 4$ matrix $\tilde{T}_{\alpha\beta}={\mc{V}_\alpha}^R{\mc{V}_\beta}^S\delta_{RS}$ of unit determinant. The eleven-dimensional metric and the four-form field of eleven-dimensional supergravity are given by
\begin{eqnarray}
d\hat{s}^2_{11}&=&\Delta^{\frac{1}{3}}ds^2_7+\frac{2}{g^2}\Delta^{-\frac{2}{3}}X^3\left[X\cos^2\xi
+X^{-4}\sin^2\xi \tilde{T}^{-1}_{\alpha\beta}\mu^\alpha\mu^\beta\right]d\xi^2\nonumber \\
& &-\frac{1}{g^2}\Delta^{-\frac{2}{3}}X^{-1}\tilde{T}^{-1}_{\alpha\beta}\sin \xi \mu^\alpha d\xi D\mu^\beta+\frac{1}{2g^2}\Delta^{-\frac{2}{3}}X^{-1}\tilde{T}^{-1}_{\alpha\beta}\cos^2\xi D\mu^\alpha D\mu^\beta\nonumber \\
& &
\end{eqnarray}
and 
\begin{eqnarray}
\hat{F}_{(4)}&=&F_{(4)}\sin\xi+\frac{1}{g}X^4\cos \xi *F_{(4)}\wedge d\xi +\frac{1}{g^3}\Delta^{-2}U\cos^5\xi d\xi \wedge \epsilon_{(3)}\nonumber \\
&
&+\frac{1}{3!g^3}\epsilon_{\alpha\beta\gamma\delta}\Delta^{-2}X^{-3}\sin\xi
\cos^4\xi \mu^\kappa
\left[5\tilde{T}^{\alpha\kappa}X^{-1}dX\right. \nonumber \\
& &\left. +D\tilde{T}^{\alpha\kappa}\right]\wedge
D\mu^\beta\wedge D\mu^\gamma \wedge D\mu^\delta+\frac{1}{2g^2}\cos\xi\epsilon_{\alpha\beta\gamma\delta}\left[\frac{1}{2}\cos\xi \sin \xi X^{-4}D\mu^\gamma\right. \nonumber \\
& &\left.-\left(X^{-4}\sin^2\xi \mu^\gamma+X^2\cos^2 \xi \tilde{T}^{\gamma\kappa}\mu^\kappa\right)d\xi\right]\wedge F^{\alpha\beta}_{(2)}\wedge D\mu^\delta\nonumber \\
& &+\frac{1}{2g^3}\epsilon_{\alpha\beta\gamma\delta}\Delta^{-2}\cos^3\xi
\mu^\kappa\mu^\lambda \left[
\cos^2\xi X^2 \tilde{T}^{\alpha\kappa}D\tilde{T}^{\beta\lambda}-\sin^2\xi X^{-3}\delta^{\beta\lambda}D\tilde{T}^{\alpha\kappa}\right.\nonumber \\
& &\left.-5\sin^2\xi \tilde{T}^{\alpha\kappa}X^{-4}\delta^{\beta\lambda}dX\right]\wedge D\mu ^\gamma\wedge D\mu^\delta \wedge d\xi
\end{eqnarray}
where $D\mu^\alpha =d\mu^\alpha+gA^{\alpha\beta}_{(1)}\mu^\beta$ and
\begin{eqnarray}
U&=&\sin^2\xi
\left(X^{-8}-X^{-3}\tilde{T}_{\alpha\alpha}\right)\nonumber \\
& &+\cos^2\xi
\mu^\alpha\mu^\beta\left(
2X^2\tilde{T}_{\alpha\gamma}\tilde{T}_{\gamma\beta}-X^2\tilde{T}_{\alpha\beta}\tilde{T}_{\gamma\gamma}
-X^{-3}\tilde{T}_{\alpha\beta}\right),\nonumber \\
\Delta &=&\cos^2\xi X\tilde{T}_{\alpha\beta}\mu^\alpha\mu^\beta+X^{-4}\sin^2\xi,\nonumber \\
\epsilon_{(3)}&=&\frac{1}{3!}\epsilon_{\alpha\beta\gamma\delta}\mu^\alpha
D\mu^\beta \wedge D\mu^\gamma \wedge D\mu^\delta\, .
\end{eqnarray}
The seven-dimensional fields and parameters are identified as follow
\begin{eqnarray}
g_2=g_1=-16h=-2g,\qquad X=e^{-\frac{\sigma}{2}}, \qquad H_{(4)}=\frac{1}{\sqrt{2}}F_{(4)}\, .
\end{eqnarray}
Relations involving the $SO(4)$ gauge fields are more complicated. Since, in this work, we do not consider the explicit embedding of solutions with non-vanishing gauge fields, we will not give them here.
 
 
\end{document}